# Monolithic integration of diverse crystalline thin films on diamond for near-junction thermal management

Tiancheng Zhao[1,2,#], Tianqi Bai[3,#], Yang He[4,#], Wenhui Xu[1,2,*,#], Xinxin Yu[5,#], Ruochen Shi[3], Zhenyu Qu[1,2], Jiaxin Liu[3], Rui Shen[5], Haodong Jiang[1], Yeliang Wang[1,2], Jiaxin Ding[1,2], Dongchen Sui[1,2], Shibin Zhang[1,2], Lei Zhu[1,2], Ailun Yi[1,2], Kai Huang[1,2], Min Zhou[1], Huarui Sun[4,*], Zhonghui Li[5], Peng Gao[3,6,*], Tiangui You[1,2], and Xin Ou[1,2,*]

[1]*State Key Laboratory of Materials for Integrated Circuits,*

*Shanghai Institute of Microsystem and Information Technology, Shanghai 200050, China*

[2]*Center of Materials Science and Optoelectronics Engineering,*

*University of Chinese Academy of Sciences, 100049 Beijing, China*

[3]*International Center for Quantum Materials, School of Physics,*

*Peking University, Beijing 100871, China*

[4]*School of Science and Ministry of Industry and Information Technology Key Laboratory of*

*Micro-Nano Optoelectronic Information System,*

*Harbin Institute of Technology, Shenzhen 518055, China*

[5]*CETC Key Laboratory of Carbon-based Electronics,*

*Nanjing Electronic Devices Institute, Nanjing 210016, China*

[6]*Tsientang Institute for Advanced Study, Hangzhou 310024, China.*

[*]*Corresponding authors:* xuwh@mail.sim.ac.cn, huarui.sun@hit.edu.cn,

p-gao@pku.edu.cn, ouxin@mail.sim.ac.cn

[#]*Authors contributed equally to this work.*

**Abstract**

The pursuit of extreme miniaturization and high power in 6G RF front-ends has cast thermal dissipation as the central challenge. Here, we have demonstrated the monolithic integration of functionally distinct single-crystal thin films, including β-$Ga_2O_3$, Si, GaN, and $LiTaO_3$, onto a single diamond substrate using a multi-step transfer printing technique. Focusing on the critical β-$Ga_2O_3$/diamond interface, we achieve an exceptional interfacial thermal conductance (ITC) of 149 MW $m^{-2}$ $K^{-1}$ through ultra-high vacuum (UHV) annealing, creating an atomically sharp interface featuring covalent bonding. Vibrational electron energy-loss spectroscopy (EELS) analysis combining with molecular dynamics (MD) simulations reveal that distinctive interfacial phonon modes at the β-$Ga_2O_3$/diamond heterointerface dominate ultrahigh ITC. We experimentally demonstrate that by improving the ITC, the thermal resistance ($R_{th}$) of a diamond-based β-$Ga_2O_3$ MOSFET is driven to a record-low value of 1.58 K mm $W^{-1}$, underscoring the critical role of interface engineering in near-junction thermal management for diamond-integrated devices. This work demonstrates a scalable, diamond-based monolithic integration platform designed to solve the near-junction thermal challenges in high-power RF front-ends.

# 1. Introduction

The evolution of 6G, low earth orbit (LEO) satellites, and radars drives RF front-ends into mm-wave/sub-THz regimes, demanding high bandwidth and power efficiency[1–3]. In these bands, higher frequencies necessitate reduced array pitch, requiring ultra-compact modules to minimize interconnect losses[4,5]. However, traditional PCB-level integration has approached its scaling limits, while emerging 2.5D/3D packaging face critical challenges concerning thermal management and mechanical reliability, specifically heat accumulation in stacked dies and stress-induced failures arising from coefficient of thermal expansion (CTE) mismatches[6,7]. Addressing these issues necessitates the direct heterogeneous integration of wide-bandgap semiconductors and diverse functional materials onto high-thermal-conductivity substrates, particularly diamond[8–13]. Such material-level integration offers a promising platform for future high-performance, miniaturized RF front-ends.

Traditional heteroepitaxy poses challenges for integrating dissimilar high-quality single-crystal thin films onto monolithic diamond substrates, namely: (i) severe lattice mismatch compromising crystalline quality; and (ii) incompatible thermal budgets during sequential growth degrading the properties of pre-grown thin films[14]. Recent breakthroughs in transfer printing[15–21], exemplified by the assembly of 2D material stacks[22,23], have established a universal framework for physically integrating dissimilar materials. However, this approach relies critically on the construction of high-quality van der Waals (vdW) interfaces between the 2D material and the substrate to enable controlled release. Translating this versatility to the heterogeneous integration of diverse functional thin-film semiconductors (tens to hundreds of nanometers) remains challenging due to the strong interfacial covalent bonding inherent to conventional epitaxial growth. To overcome this limitation, the X-on-Insulator (XOI) architecture emerges as a tailor-made framework for realizing diverse functional thin films on SiO$_2$/Si substrates. Its buried dielectric structure naturally accommodates sacrificial layer-assisted exfoliation, paving the way for the material-level monolithic integration.

Our prior work first demonstrated successful integration of arrayed β-Ga$_2$O$_3$ single-crystalline thin films onto 1-inch diamond substrate[24]. Despite this structural success, the initial demonstration was confined to single-material transfer. Achieving multi-material integration on diamond is impeded by the difficulties in maintaining thin-film integrity, modulating competitive interfacial adhesion and attaining large-area placement precision. Furthermore, as the near-junction thermal dissipation is dominated by the interfacial thermal conductance (ITC)[25], the quality of heterointerfaces is particularly critical for unlocking the full potential of diamond. Therefore, advanced interface engineering is vital for next-generation RF front-ends, underscored by the industry's growing pursuit of diamond-integrated architectures to efficiently manage the extreme heat flux of densely packed active devices.

In this work, we firstly demonstrate monolithic integration of arrayed β-Ga$_2$O$_3$, Si, GaN, and LiTaO$_3$ single-crystal thin films on a 1-inch diamond substrate via a multi-step transfer printing technique. We construct high-quality β-Ga$_2$O$_3$/diamond heterogeneous interfaces with ultra-high ITC via ultra-high vacuum (UHV) annealing or a low-temperature interlayer-assisted interface reinforcement method, respectively. The dominant mechanism of interfacial thermal transport at the β-Ga$_2$O$_3$/diamond interface is elucidated using coupled molecular dynamics (MD) simulations and electron energy-loss spectroscopy (EELS) analysis. The ITC enhancement strategy enables a diamond-based β-Ga$_2$O$_3$ MOSFET with a record-low $R_{th}$ of 1.58 K mm W$^{-1}$. This work offers a scalable platform for monolithic RF front-end integration on diamond, enabling advanced near-junction thermal management.

## 2. Results and Discussion

### 2.1. Monolithic Integration of Diverse Functional Thin Films on Diamond

Monolithic integration of single-crystal thin films on diamond has been achieved, utilizing β-$Ga_2O_3$ for Power Amplifier (PA), Si for Logic, GaN for RF Switch and Low Noise Amplifier (LNA), and $LiTaO_3$ for Filter. This platform addresses the material-level demands of high-density RF front-end modules. With the adoption of diamond substrates for efficient thermal management, the high breakdown field of β-$Ga_2O_3$ can be fully leveraged to realize power amplifiers with record-breaking power handling capabilities. Si remains the preferred choice for control functions, offering a cost-efficient solution for processing digital logic, managing intricate timing sequences, and generating high-precision bandgap voltage references. GaN serves as a dual-role component, functioning both as a low-loss power switch and a highly robust LNA. Moreover, the $LiTaO_3$-on-diamond heterostructure facilitates the maximum spectral purity required for next-generation communication standards. Fig. 1a illustrates the fabrication process. Patterning the thin film mitigates the severe stress-induced damage (crack or fracture) encountered during full large-area thin film transfer process, while introducing regular channels for sacrificial layer release. We realized the high-integrity, large-scale retrieval of thin-film arrays by tailoring the undercutting conditions for various XOI materials. And four distinct thin-film microarrays are sequentially integrated at their designated sites. Fig. 1b displays the optical images of a 1-inch diamond substrate captured after each successive transfer step. Alignment marks, appearing as gray spots, are lithographically patterned on the diamond substrate to serve as positional references for the sequential thin-film transfer. And the inset optical micrographs reveal that the as-transfer-printed (As-TP) thin films maintain structural integrity, exhibiting no visible cracks. Raman spectroscopy (Fig. S1) corroborates the successful integration of four thin films on diamond, as the spectra clearly resolve the characteristic vibrational modes of each thin film concurrently with the signature 1332 cm$^{-1}$ diamond peak[26]. AFM measurements across a 5 × 5 μm$^2$ scan area reveal that the root-mean-square (RMS) surface roughness of the four thin films is less than 1 nm (Fig. S2), meeting the fundamental criteria for device fabrication. Fig. 1c characterizes the microstructure of these As-TP thin films on diamond. The well-ordered lattice in the transmission electron microscope (TEM) images and the high-resolved atomic columns in the high-angle annular dark-field scanning TEM (HAADF-STEM) micrographs indicate that the β-$Ga_2O_3$, Si, GaN, and $LiTaO_3$ thin films retain excellent single-crystallinity after multi-step transfer-printing. This confirms the feasibility of XOI-enabled transfer printing for advanced, highly integrated RF front-ends, concurrently increasing functionality density, reducing module area, and suppressing parasitic capacitance[27].

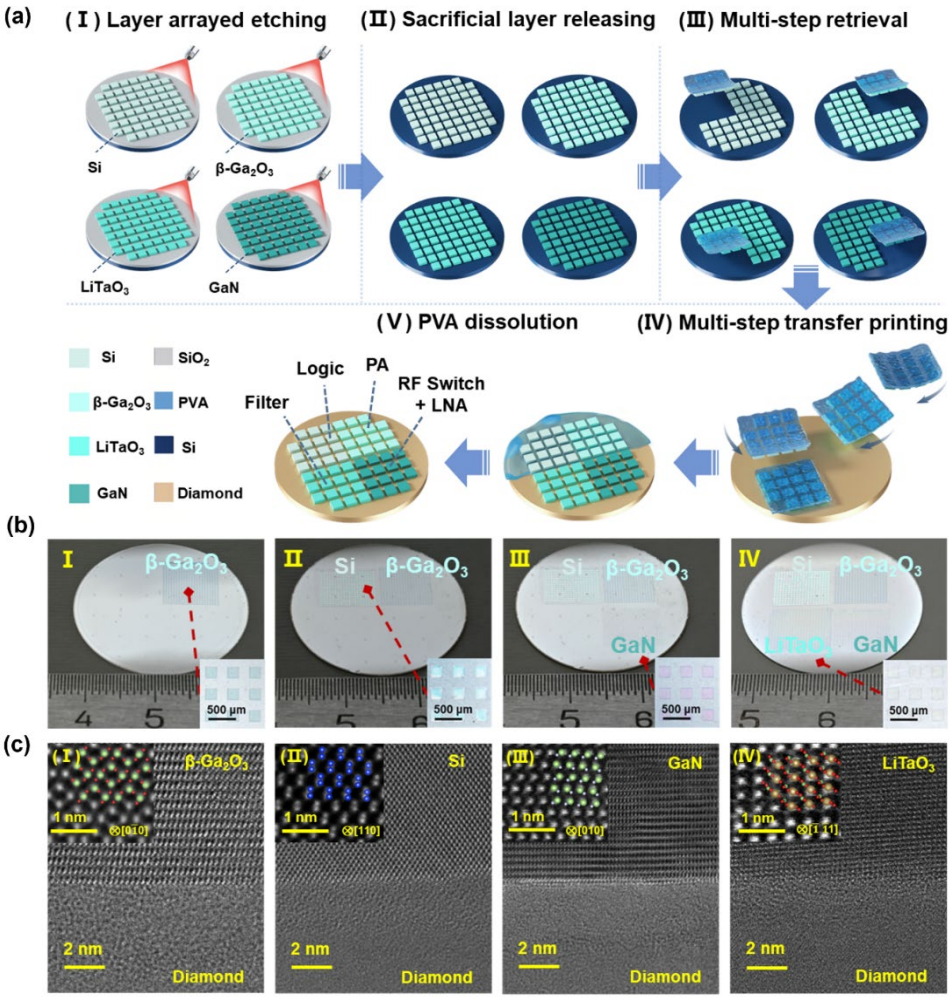

**Fig. 1. Monolithic Integration of functionally distinct single-crystal thin films on diamond.** (a) Process flow for integration of four thin films (β-$Ga_2O_3$, Si, GaN, and $LiTaO_3$) on diamond substrates by XOI-enabled transfer printing. (b) Photographs taken after each transfer step of arrayed thin films on diamond. The gray spots correspond to the alignment marks patterned on the diamond substrate to assist the multi-step transfer process. Insets: Optical micrographs of As-TP thin films (scale bar: 500 μm). (c) Cross-sectional TEM images of thin-film/diamond heterostructures and corresponding atomic-resolution HAADF-STEM characterization of thin films. Scale bars represent 2 nm and 1 nm, respectively.

## 2.2. Interface Engineering of Diamond-Based Heterostructure

Thermal management is fundamental to the advancement of next-generation high-performance hetero-integrated electronics[28,29]. Particularly for diamond-based architecture, given the extremely low thermal resistance of diamond substrate in heat transfer path, it is imperative to improve the ITC via engineering interface thermal transport properties, ensuring efficient heat extraction from the active junction[30]. Weak bonding and phonon mismatch at heterointerface can give rise to low ITC[31,32]. Primary interface engineering methods include thermal annealing to modify the interfacial microstructure[33–35] and the insertion of an interlayer to bridge the vibrational gap between two materials with mismatched phonon transport properties[36–41].

As $\beta$-$Ga_2O_3$ possesses the lowest thermal conductivity and thus presents the most significant challenge for heat dissipation among the integrated thin films, we systematically engineered a series of $\beta$-$Ga_2O_3$/diamond heterointerfaces to maximize the ITC. A novel UHV annealing process is employed to promote the formation of an atomically flat interface by maintaining an ultra-clean thermal environment[42,43]. The bright-field STEM (BF-STEM) images (Fig. 2a) compare the interface structures of the As-TP and UHV-annealed (UHV-Ann) $\beta$-$Ga_2O_3$/diamond interface, along with those incorporating different interlayers (ILs), including $SiO_2$-IL, $SiN_x$-IL, and AlN-IL. The As-TP interface exhibits an amorphous transition layer ~2 nm thick (Fig. 2a, I), whereas the UHV-Ann interface is well-defined and atomically sharp (Fig. 2a, II). And the interlayer thickness for samples featuring distinct inserted materials is maintained at ~3 nm (Fig. 2a, III-V). The thermal transport properties are evaluated using transient thermoreflectance (TTR). Of the heterointerfaces investigated, the UHV-Ann sample exhibits the most efficient heat dissipation (Fig. S3). The ITC is extracted by performing a multi-parameter fit of the TTR signal using a transient heat transfer model combined with a quantum genetic algorithm[44,45]. As depicted in Fig. 2b, the UHV-Ann sample demonstrates the highest average ITC (ITC$_{avg}$) of 118 MW m$^{-2}$ K$^{-1}$, which is enhanced by a factor of ~4 compared to its As-TP sample (29 MW m$^{-2}$ K$^{-1}$). In comparison to the As-TP interface, the heterointerfaces employing distinct interlayers exhibit different levels of improvement in ITC, which is indicative of a phonon bridging mechanism, albeit less pronounced than the UHV-Ann interface (detailed values are listed in Table S1). By tailoring the interfacial atomic structure, we effectively bridge the vast phonon mismatch between diamond and β-Ga$_2$O$_3$, as discussed later. ITC mapping is employed to provide a comprehensive comparative assessment of the two interface engineering strategies (Fig. S4), and the UHV-Ann sample exhibits a superior uniformity (calculated as defined in Supplementary Note 3) of ~80%.

A further key advantage of interface modification in this work is the enhancement of interfacial adhesion. The As-TP thin films bonded to the substrate primarily by weak van der Waals forces, are susceptible to detachment during treatment with solutions like piranha. Conversely, UHV annealing constructs a chemically robust interface ensuring its compatibility with subsequent device fabrication processes. To quantify the change in interfacial strength, the As-TP and UHV-Ann samples are subjected to scratch adhesion test[46] (Fig. S5-6). The critical load corresponding to large-scale thin film delamination increased from ~0.3 N for the As-TP sample to ~0.8 N for the UHV-Ann, revealing a considerable improvement in bonding strength (Fig. 2c). The bond energy of the UHV-Ann heterointerface is estimated to be ~186 kJ mol$^{-1}$ (see Supporting Information, Section 4 for detailed calculations). This value suggests covalent characters at the interface, significantly exceeding typical Van der Waals interactions[47]. The formation of an intermixed region of C, O and Ga atoms in the UHV-Ann heterointerface, confirmed by EELS core-loss spectra in Fig. 2d, serves as compelling evidence for the enhanced bonding strength. Consequently, the measured ITC of the best-performing UHV-Ann interface (149 MW m$^{-2}$ K$^{-1}$ at 300 K) and its subsequent temperature-dependent increase, closely follows the theoretical predictions of a structurally abrupt interface derived from MD simulations, far surpassing the values obtained from the As-TP interface (Fig. 2e). In addition, the introduction of an interlayer can also effectively improve bonding strength. The interlayer-assist interface modulation approach without resorting to high temperatures offers a low-temperature-compatible strategy for monolithic integration.

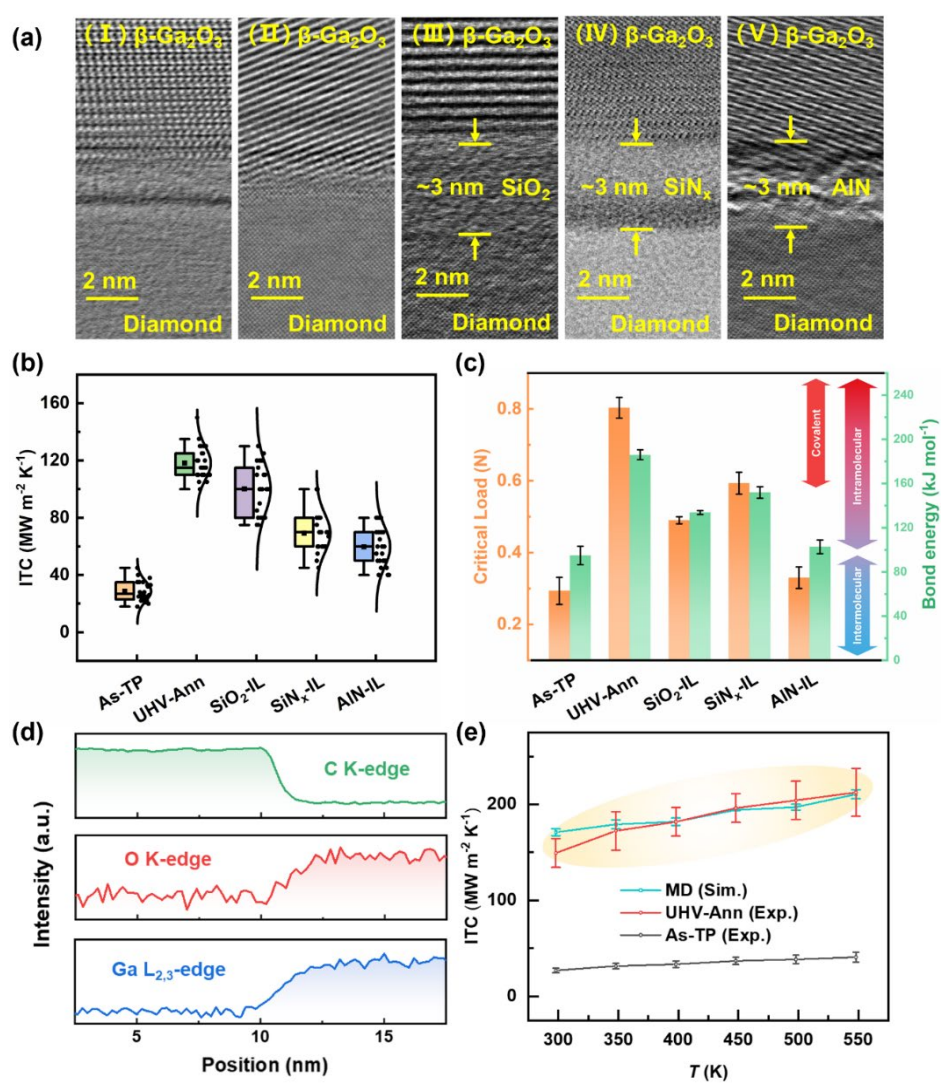

**Fig. 2. Interface engineering and characterization of diamond-based heterostructure for thermal management.** (a) BF-STEM images of the (I) As-TP, (II) UHV-Ann, (III) SiO$_2$-IL,

(IV) SiN$_x$-IL, and (V) AlN-IL β-Ga$_2$O$_3$/diamond heterointerfaces (scale bar: 2 nm). (b) Comparison of ITC across the As-TP, UHV-Ann, and interlayer-modified β-Ga$_2$O$_3$/diamond heterointerfaces. ITC statistics are derived from 25 spatially distributed points on each sample. (c) The critical delamination loads and estimated bond energies for the corresponding interfaces. Error bars denote the standard deviation of three independent measurements. (d) Core-loss EELS at UHV-Ann interface. Distinct mixed arrangement of C, O and Ga atoms is observed at interface. (e) Temperature-dependent ITC calculated by MD in comparison with experimental data from TTR.

## 2.3. Interfacial Thermal Transport Mechanism

To elucidate the physical mechanisms underlying the enhanced thermal transport, we probed the phonon modes at the heterointerface using vibrational EELS. Prior to UHV annealing (Fig. 3a), virtually no interfacial phonon modes are observed, and the bulk phonon peaks at the β-Ga$_2$O$_3$/diamond interface present a gradual transition. In contrast, highly localized interfacial phonon modes emerge at the β-Ga$_2$O$_3$/diamond interface after UHV annealing (Fig. 3b), confined to a region of only 1-2 nm. The calculated EELS line profile shown in Fig. S7 reproduces the experimental features remarkably well. After subtracting the bulk contributions from interfacial spectrum, the extracted residual reflects the change in the phonon densities of states (PDOS) at the interface (Fig. 3c). Peaks in the residual are attributed to new modes localized at the interface (localized modes), whereas dips indicate a suppression of modes (isolated modes) that are present in the bulk[48]. The interface modes are attributed to the unique bonding configuration at the interface, a zone of atomic intermixing involving C, O and Ga species, distinct from the bulk environments of β-Ga$_2$O$_3$ and diamond. Fig. 3d illustrates the spatial distribution of the interface modes, revealing that both localized and isolated modes are tightly confined to the interface vicinity. Localized modes are observed at ~29, ~64, and ~99 meV, while isolated modes appear at ~43 and ~86 meV. The specific contributions of each phonon mode to total ITC are validated by MD simulations (Fig. S8) of the spectrally resolved heat flux, as shown in Fig. 3e. The calculations reveal that phonon modes below 100 meV contribute over 86.3% of the total ITC, dominated by two main peaks at ~28 and ~55 meV. This theoretical prediction aligns well with the experimentally detected localized modes at ~29 and ~64 meV, which originates from specific interface structure and strongly correlates with phonon modes on both sides to facilitate cross-interface energy transfer[49,50]. The two modes visualized as phonon eigenvectors in Fig. 3f exhibits prominent interfacial character, serving as vital phonon bridges to mediate elastic and inelastic scattering across the interface[51,52].

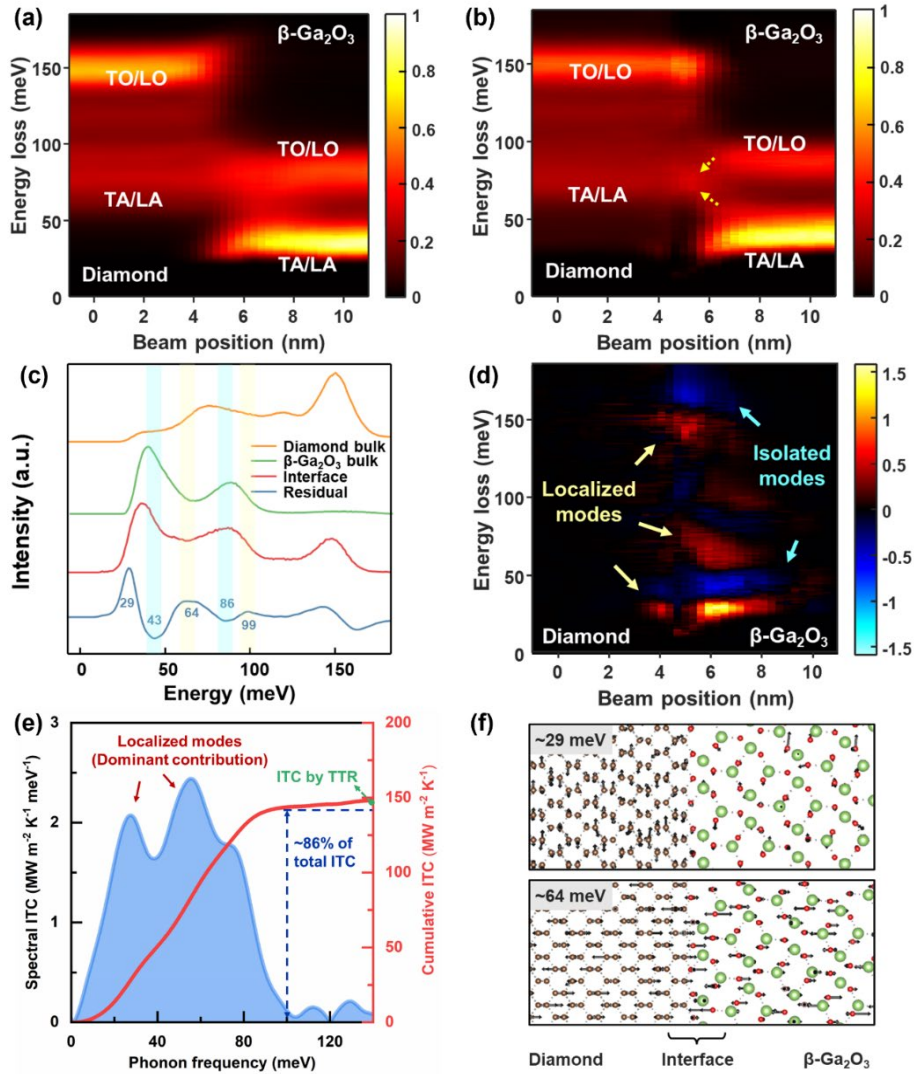

**Fig. 3. Mechanisms underlying enhanced thermal transport.** The measured EELS line profile with fitted phonon peak position of five typical phonon branches (dotted line) across the

(a) As-TP and (b) UHV-Ann β-Ga$_2$O$_3$/diamond interface. (c) The EELS acquired in diamond, in β-Ga$_2$O$_3$ and at the interface, with the fitting residual indicating the PDOS of the interface phonon modes excluding bulk contributions. (d) Line profile of the residual across the interface, where red and blue spots correspond to the localized and isolated modes, respectively. (e) The spectral and cumulative ITC at β-Ga$_2$O$_3$/diamond interface from MD. (f) The calculated phonon eigenvectors with arrows illustrating the vibration amplitude of each atom.

## 2.4. Device Demonstrations

To experimentally validate the critical contribution of high ITC to the device's thermal performance, RF MOSFETs are fabricated on UHV-Ann β-Ga$_2$O$_3$/diamond heterostructure (Fig.4a). Fig. 4b plots the device junction temperature rise ($\Delta T_j$), measured by an IR camera, as a function of applied DC power ($P_{DC}$) for identical β-Ga$_2$O$_3$/diamond structures featuring different ITC values. As the ITC$_{avg}$ increases from 46 MW m$^{-2}$ K$^{-1}$ to 118 MW m$^{-2}$ K$^{-1}$, the device thermal resistance ($R_{th}$) decreases from 5.52 K mm W$^{-1}$ to 1.58 K mm W$^{-1}$. This corresponds to a 71% reduction, confirming the dominance of ITC in the overall $R_{th}$ and underscoring the vital role of interfacial engineering for heat dissipation in diamond-integrated devices. The extremely low $R_{th}$ represents approximately only 1/40 of the $R_{th}$ extracted in β-Ga$_2$O$_3$ bulk RF devices (62.4 K mm W$^{-1}$)[24], thereby successfully overcoming the heat dissipation bottleneck caused by the low thermal conductivity of β-Ga$_2$O$_3$. Notably, the inset depicts a peak temperature rise as low as ~18 K at the power of 1 W, providing a substantial thermal safety margin against thermal degradation. As evidenced by the 2D heat maps (Fig. 4c) simulated via finite element method (FEM), the improved ITC leads to a more uniform temperature distribution, effectively suppressing potential failure-inducing hotspot. Fig. 4d[24,37,53–55] benchmarks the $R_{th}$ of various diamond-based devices as a function of the ITC of their heterostructures. Advanced interface engineering yields an outstanding ITC in β-Ga$_2$O$_3$/diamond heterostructure, in turn driving the device $R_{th}$ to a leading value. Furthermore, the β-Ga$_2$O$_3$/diamond heterogeneous device fabricated in this work demonstrates the lowest $R_{th}$ while sustaining an ultra-high power density (~10 W mm$^{-1}$), further verifying the effectiveness of the near-junction thermal management strategy (Fig.4e)[24,56]. This achieved ultra-low $R_{th}$ is key to enabling high power density, linearity, and reliability in next-generation RF front-ends.

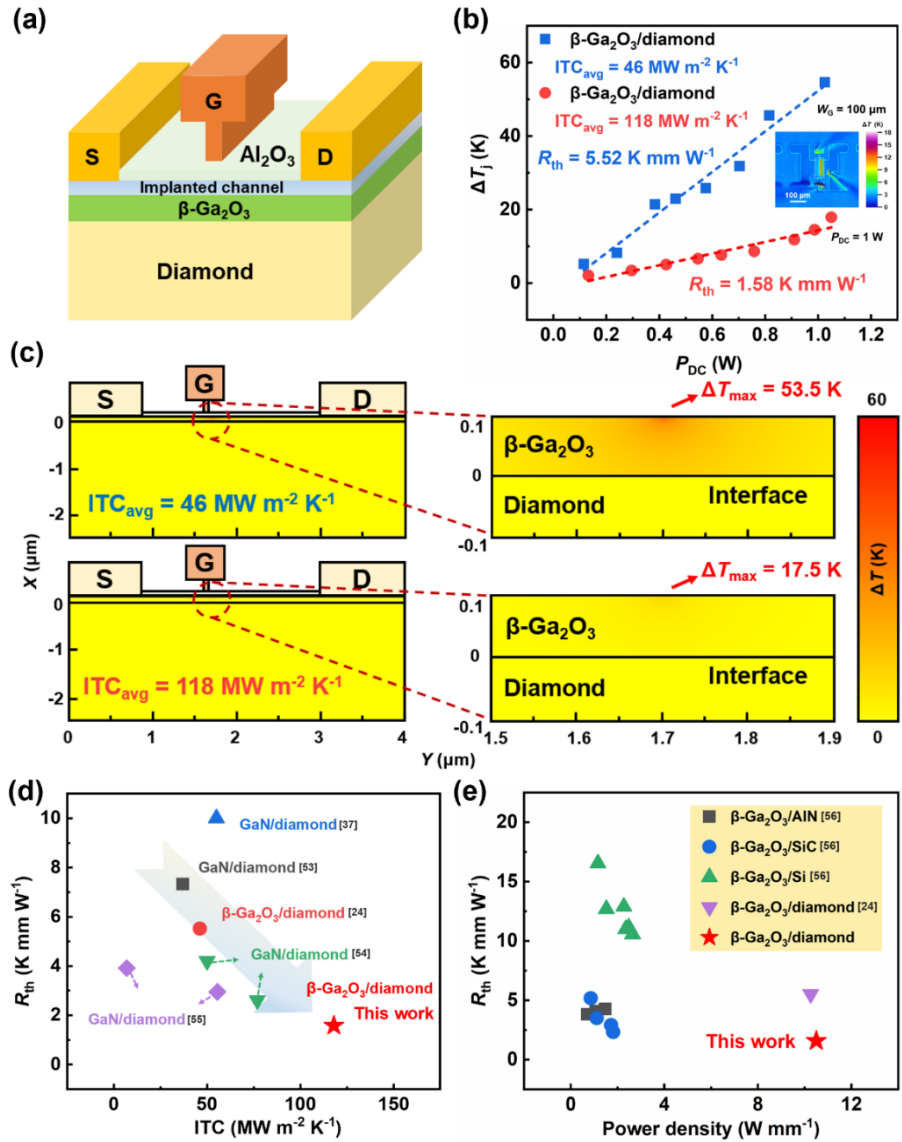

**Fig. 4. Device demonstrations.** (a) Schematic diagram of the β-Ga$_2$O$_3$ RF MOSFET on diamond substrate with a T-gate. (b) The $\Delta T_j$ of the identical β-Ga$_2$O$_3$/diamond heterostructure with different ITC$_{avg}$ at various applied $P_{DC}$. (c) FEM-simulated 2D heat maps of the MOSFET cross-section. (d) Benchmark of $R_{th}$ vs. ITC for diamond-based heterogeneous integrated

device[24,37,53–55]. (e) $R_{th}$ versus applied power density for β-Ga$_2$O$_3$ RF MOSFETs on Si, SiC, AlN and diamond substrates[24,56].

## 3. Conclusion

We have demonstrated a robust diamond-based monolithic integration platform incorporating functionally distinct single-crystal thin films (β-Ga$_2$O$_3$, Si, GaN, and LiTaO$_3$) via XOI-enabled transfer printing technology. The precise, high-integrity transfer of large-area thin-film arrays is achieved by optimizing the sacrificial layer release parameters for diverse XOI stacks and employing lithography-assisted alignment methods. The AFM characterization confirms that the surface morphology remains smooth, satisfying the strict roughness requirements for high-yield device processing. Moreover, TEM investigations reveal that the films retain their pristine single-crystal nature, devoid of threading dislocations or defects typical of direct hetero-epitaxy. These structural attributes establish a solid foundation for the realization of the theoretical performance limits of the integrated materials.

Since the lowest-thermal-conductivity β-Ga$_2$O$_3$ represents the critical limiting factor for heat dissipation, we focused on optimizing β-Ga$_2$O$_3$/diamond heterointerfaces to boost the ITC. Our results highlight two distinct yet complementary strategies for interface engineering. By employing UHV annealing, an atomically sharp, covalently bonded β-Ga$_2$O$_3$/diamond interface is realized. The optimized interface structure facilitates efficient phonon transport, resulting in a superior ITC up to 149 MW m$^{-2}$ K$^{-1}$. Conversely, the interlayer-assisted modulation provides an effective low-thermal-budget alternative for temperature-sensitive applications. While this approach may not reach the peak value in ITC, it ensures reliable thermal management without compromising the integrity of the integrated thin films.

Experimental EELS results, corroborated by MD simulations, identifies unique interfacial phonon modes at ~29 and ~64 meV as the dominant contributor to the ultra-high ITC of UHV-Ann interface. These interface modes are attributed to the unique bonding configuration at the interface, characterized by a confined zone of atomic mixing involving C, O, and Ga species. Crucially, the two modes acts as phonon bridges, effectively mediating the vibrational mismatch between the bulk β-$Ga_2O_3$ and diamond, thereby facilitating efficient heat transfer. Benefiting from the engineered interface and diamond's superior heat spreading, the $R_{th}$ of β-$Ga_2O_3$ RF MOSFET fabricated on the diamond platform is reduced to a record-low value of 1.58 K mm $W^{-1}$, thereby removing the barriers to the deployment of high-power RF front-ends. This work provides a viable technological pathway for the development of next-generation, high heat flux density, and highly integrated systems-on-diamond.

## 4. Methods

*Sample preparation*: For in-plane multi-material heterogeneous integration, the top-device-layers of XOI wafers (β-$Ga_2O_3$/$SiO_2$/Si, Si/$SiO_2$/Si, GaN/$SiO_2$/Si and $LiTaO_3$/$SiO_2$/Si) fabricated by ion-cutting technique[57–59] are patterned into 200 μm×200 μm square thin-film microarrays via inductively coupled plasma reactive ion etching (ICP-RIE) (Etching recipes detailed in Table S2). Then the $SiO_2$ layers are removed by wet etching, and high-fidelity release of thin films is achieved by tuning the etchant concentration and etching time (Etching recipes detailed in Table S3). Using a custom-built transfer system (Shanghai ONWAY Technology Co., Ltd), thin-film microarrays are reliably retrieved from Si substrates with prepared polyvinyl alcohol (PVA) films serving as a versatile stamp. And the integrated CCD microscopic imaging module allows for the site-specific transfer of thin films onto photolithographically defined locations on the diamond substrate. Finally, the PVA films are dissolved in ultrapure water, enabling the heterogeneous integration of dissimilar materials onto the diamond substrate.

*Device Fabrication*: Firstly, the arrayed β-$Ga_2O_3$ thin film on diamond substrate is subjected to Si ion implantation at an acceleration energy of 15 keV and a target dose of 3×$10^{13}$ $cm^{-2}$ to form a shallowly implanted channel. Subsequently, the implanted Si ions are activated by rapid thermal annealing (RTA) at 950°C in $N_2$ ambient for 2 minutes. The source and drain electrode patterns are defined by photolithography, followed by the deposition of Ti/Au (20/200 nm) metal stacks via electron beam evaporation (EBE). A low ohmic contact resistance is obtained following an RTA step at 470°C in $N_2$ for 3 minutes. A high-quality 15 nm $Al_2O_3$ gate dielectric is grown via ALD at 350°C. The Ni/Au (20/500 nm) T-shaped gate electrodes are defined using a bi-layer resist process in combination with electron beam lithography (EBL).

*TTR*: The TTR optical path is based on a pump–probe configuration. A 355 nm pulsed pump laser is aligned coaxially with a 532 nm continuous-wave probe laser using a dichroic mirror. Both beams are then focused onto the sample surface. Critically, the pump spot diameter is made significantly larger than that of the probe beam. This design ensures that the transient thermal response approximates quasi-one-dimensional longitudinal heat conduction, enhancing measurement accuracy. The thermophysical properties of the heterostructures are extracted by fitting the TTR signals using a quantum genetic algorithm (QGA). Leveraging the principles of quantum superposition, the QGA possesses superior global optimization capabilities, making it particularly suited for non-convex optimization problems inherent in multiparameter extraction. Integrating the QGA with the TTR technique minimizes solution randomness and enables the precise, simultaneous determination of multiple thermal parameters.

*Hardware*: The STEM-EELS data is acquired on a Nion U-HERMES200 microscope equipped with both a monochromator and aberration correctors. The EELS datasets are acquired with 60 kV beam energy and 35 mrad convergence semi-angle and 25 mrad collection semi-angle. The typical energy resolution of acquired data under these conditions is 19 meV. Off-axis[60] collection is used to mitigate the delocalization effect due to the dipole scattering. The scanning step size is 1/3 and 1/3.5 nm for the annealed and unannealed sample, and the typical spatial resolution is ~0.3 nm. The typical dwell times of each pixel is 1600-2000 ms.

*EELS Data processing*: All the acquired vibrational spectra are processed by using the custom-written MATLAB code and Gatan Microscopy Suite. More specifically, the EEL spectra are first aligned. Subsequently, the 3D filtering (BM3D) algorithms are applied for removing the Gaussian noise[61]. The background arising from both the tail of the zero-loss peak (ZLP) and the non-characteristic phonon losses is fitted by employing the modified Pearson-VII function with two fitting windows and then subtracted in order to obtain the vibrational signal[62]. The Lucy-Richardson algorithm is then employed to ameliorate the broadening effect induced by the finite energy resolution, taking the elastic ZLP as the point spread function. The spectra are summed along the direction parallel to the interface for obtaining line-scan data with a good signal-to-noise ratio.

*Simulations*: To calculate spectral ITC, the atomic simulation is performed using Large-scale Atomic/Molecular Massively Parallel Simulator (LAMMPS) and Quantum ESPRESSO (QE), providing detailed insights for the experimental data. Non-equilibrium molecular dynamics (NEMD) is applied for ITC calculation and spectral analysis. Especially, the core-shell model is chosen for the β-$Ga_2O_3$ with the Born-Mayer-Huggins-type potential between the cores, the harmonic potential between the core-shell in the same atom as well as the Coulomb interaction with the damped shifted force. The Tersoff potential is used for the diamond whereas the Lennard-Jones potential is employed for the interface interaction.

To calculate phonon eigenvectors, the MD simulations are carried out by Graphics Processing Unit Molecular Dynamics (GPUMD), using the trained neuro-evolution machine learning potential (NEP) model. The training dataset is generated using the ab initio MD module of the ABACUS software package[63–65]. We constructed the diamond (100) / β-$Ga_2O_3$ (-201) interface, incorporating three distinct termination structures that include both Ga-C and Ga-O interfaces. Trajectories are collected at temperatures of 100 K, 300 K, 900 K, and 1500 K, respectively. In total, we generated 600 atomic configurations of the interface systems. These are combined with 300 configurations each of bulk diamond and bulk β-Ga$_2$O$_3$ generated under similar conditions to form the complete dataset. The dataset is subsequently split, with 80% allocated for training and 20% reserved for testing. Periodic boundary conditions are applied in all directions, and the time step is set as 1 fs. The phonon eigenvectors are computed in a supercell measuring 15.0 Å × 15.0 Å × 72.5 Å, using the NEP potential generated as described above. The calculation of the PDOS is performed in a larger supercell with dimensions of 30.2 Å × 30.7 Å × 198.8 Å. The procedure involved first equilibrating the system in the NPT ensemble at zero pressure and 300 K for 30 ps. This is followed by a 30-ps simulation in the NVT ensemble at 300 K. Finally, a 500-ps production run is conducted in the NVE ensemble to compute the phonon DOS, which is obtained via the mass-weighted velocity autocorrelation (VAC) function.

To investigate the thermal behavior of β-Ga$_2$O$_3$ RF MOSFET on diamond substrate, a finite element model of is conducted. The geometric model is simplified based on the experimental specimen. The experimentally determined power is set as the heat source in the thermal simulation of the device. The bottom of heat sink is fixed at room temperature (293.15 K), and the remaining surfaces are subject to a natural convection with a heat transfer coefficient of 5 W m$^{-2}$ K$^{-1}$. A mesh convergence study is performed to ensure that the results are independent of the mesh size.

**Acknowledgements**

This work was supported by the Natural Science Foundation of China (Grant Nos: 62293521, 62404236, 12404192), the Science and Technology Commission of Shanghai Municipality (Grant No: 24DP1500300), the Shenzhen Science and Technology Program (Grant No: JCYJ20250604145537048) and the Open Research Fund of State Key Laboratory of Materials for Integrated Circuits (Grant No: SKLIC-K2024-04). And the authors acknowledge High-performance Computing Platform of Peking University for providing computational resources.

**Data Availability Statement**

The data that support the findings of this study are available from the corresponding author upon reasonable request.

# Supporting Information

## 1. Raman spectroscopy

Fig. S1 displays the Raman spectra of four thin films on diamond substrates. All spectra exhibit a dominant, sharp peak at 1332 cm$^{-1}$, which is the characteristic $T_{2g}$ mode of the diamond substrate[1]. Distinct signature Raman modes are identified for each thin film: the $A_g(3)$ mode at 200.5 cm$^{-1}$ for the β-Ga$_2$O$_3$[2]; the $T_{2g}$ mode at 520.7 cm$^{-1}$ for the Si[3]; the $E_2$(high) mode at 569 cm$^{-1}$ for the GaN[4]; and the $E(TO_1)$ and $A_1(TO_4)$ modes at 143 cm$^{-1}$ and 600 cm$^{-1}$ for the LiTaO$_3$[5].

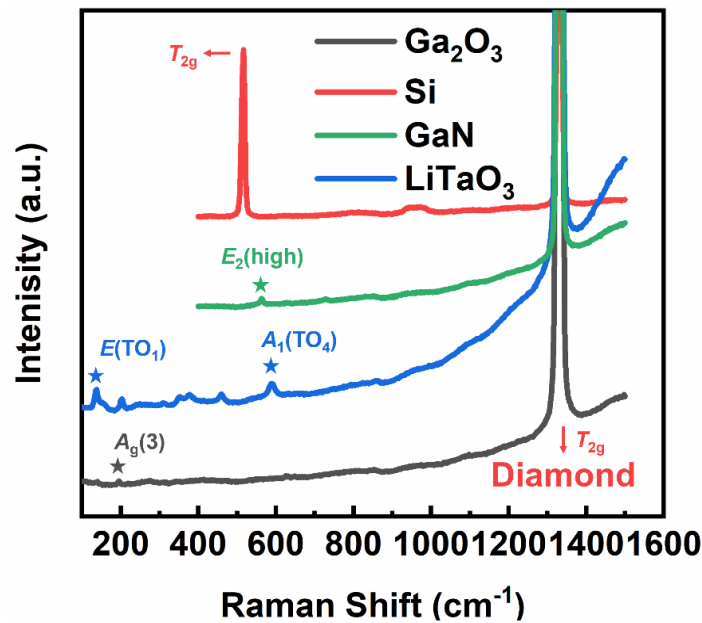

**Fig. S1.** Raman spectra identifying characteristic peaks for each functional membrane.

## 2. AFM

Fig. S2 exhibits the surface morphology of thin films after transfer printing ( 5 μm × 5 μm scan). The RMS roughness values of four films are all below 1 nm, which satisfies the basic requirements for device fabrication.

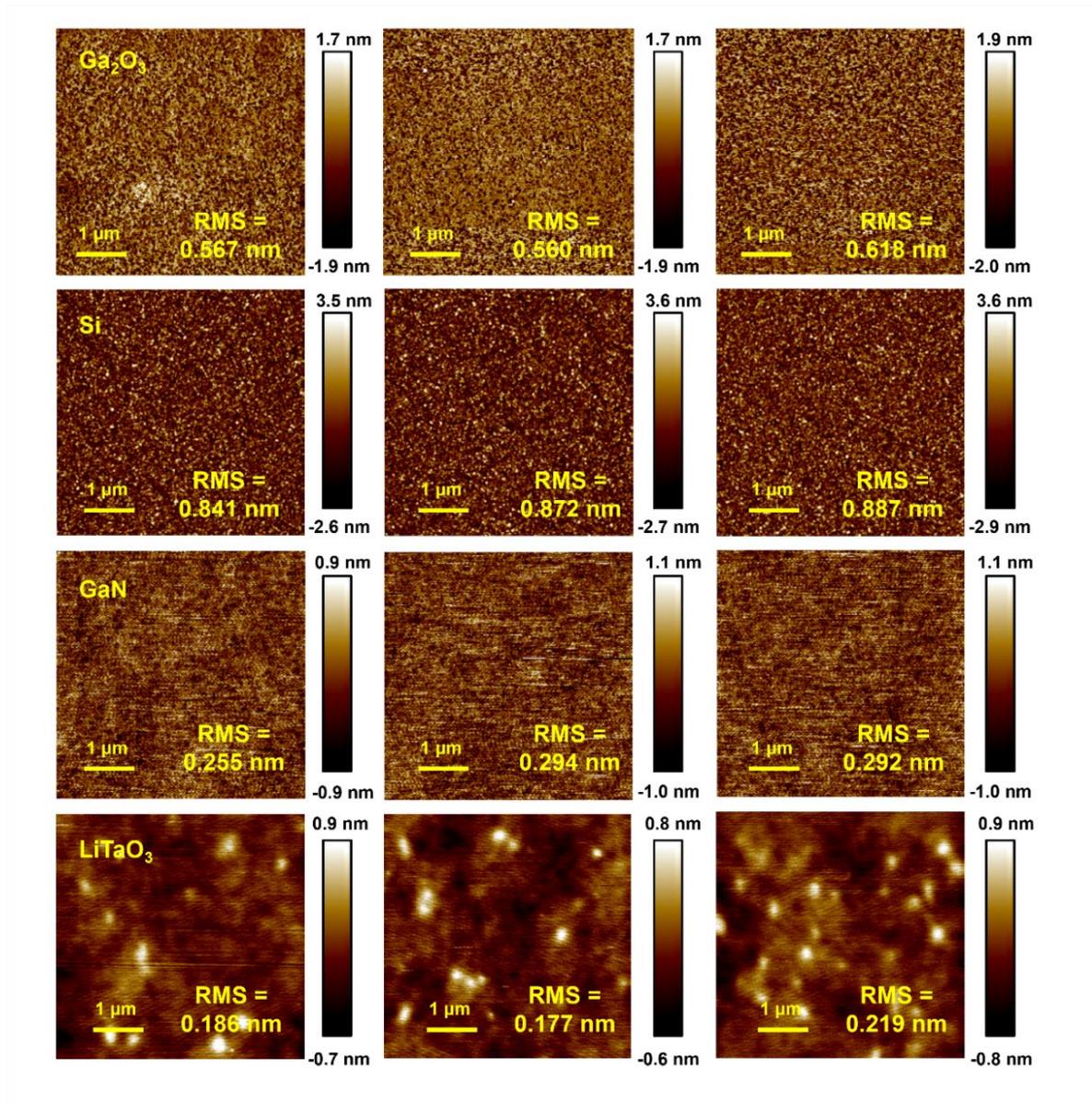

**Fig. S2.** AFM image of transfer-printed thin film on diamond substrate (scale bar: 1 μm).

## 3. Transient thermoreflectance (TTR)

Fig. S3a presents TTR curves demonstrating significantly enhanced heat dissipation rates at the UHV-Ann β-$Ga_2O_3$/diamond interface compared to the As-TP heterointerface. The sensitivity curves in Fig. S3b show a high sensitivity to ITC, indicating that the ITC significantly affects the thermal transport of β-$Ga_2O_3$ on diamond.

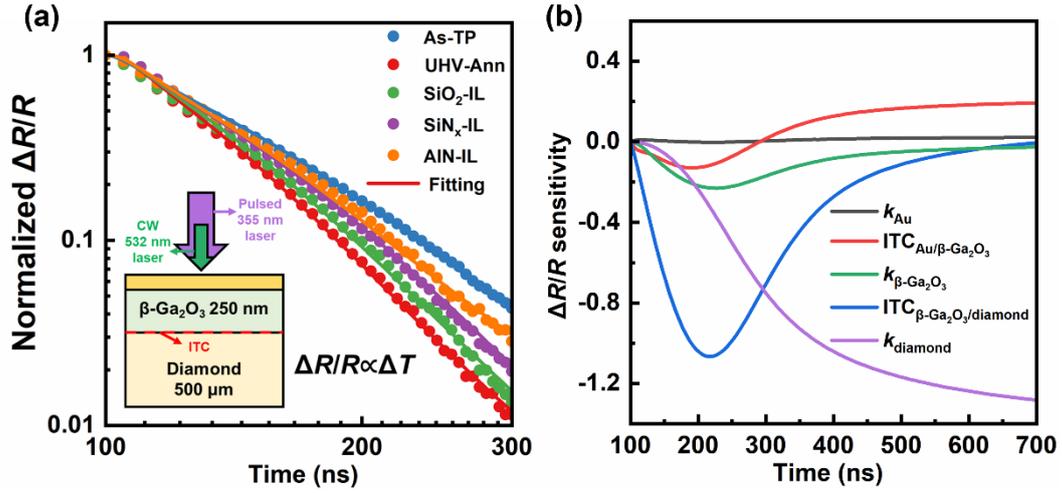

**Fig. S3.** (a) Normalized TTR curves for different β-Ga$_2$O$_3$/diamond interface configurations. (b) The sensitivity curves for β-Ga$_2$O$_3$/diamond in each input extracted parameter ($k_{Au}$, ITC$_{Au/GaO}$, $k_{GaO}$, ITC$_{GaO/Dia.}$, $k_{Dia}$) calculated by the transient heat transfer model.

The thermal conductivity of β-Ga$_2$O$_3$ thin film and ITC of each structure is extracted by fitting TTR curves with a transient heat transfer model combined with a quantum genetic algorithm[6,7]. As detailed in Table S1, UHV-Ann yields no significant improvement for the ITC of three samples with an interlayer, in contrast to the notable enhancement seen in the directly bonded samples. This suggests that the low-temperature strategy introducing an interlayer presents a suitable alternative in applications with constrained thermal budgets.

**Table S1.** TTR results.

| Sample | | Thermal conductivity of β-Ga$_2$O$_3$ thin film (W/m·K) | ITC (MW/m$^2$·K) |
| --- | --- | --- | --- |
| β-Ga$_2$O$_3$/diamond | As-TP | 8.1 ± 1 | 29 ± 7 |
| | UHV-Ann | 9.5 ± 1.5 | 118 ± 11 |
| β-Ga$_2$O$_3$/SiO$_2$/diamond | As-TP | 8.2 ± 1 | 100 ± 17 |

| | UHV-Ann | 9.3 ± 1.5 | 113 ± 12 |
| | As-TP | 7.9 ± 1 | 69 ± 13 |
| β-Ga$_2$O$_3$/SiN$_x$/diamond | UHV-Ann | 9.4 ± 1.5 | 88 ± 24 |
| | As-TP | 8.2 ± 1 | 60 ± 14 |
| β-Ga$_2$O$_3$/AlN/diamond | UHV-Ann | 9.3 ± 1.5 | 62 ± 8 |

To systematically assess the efficacy of the two competing strategies, the ITC mapping is implemented (Fig. S4). The UHV-Ann β-Ga$_2$O$_3$/diamond heterostructure exhibits higher uniformity than all the As-TP heterostructures with an interlayer. The uniformity is defined by the following equation:

$$\text{Uniformity} = \left(1 - \frac{\text{Maximum} - \text{Minimum}}{2 \times \text{Mean}}\right) \times 100\%$$

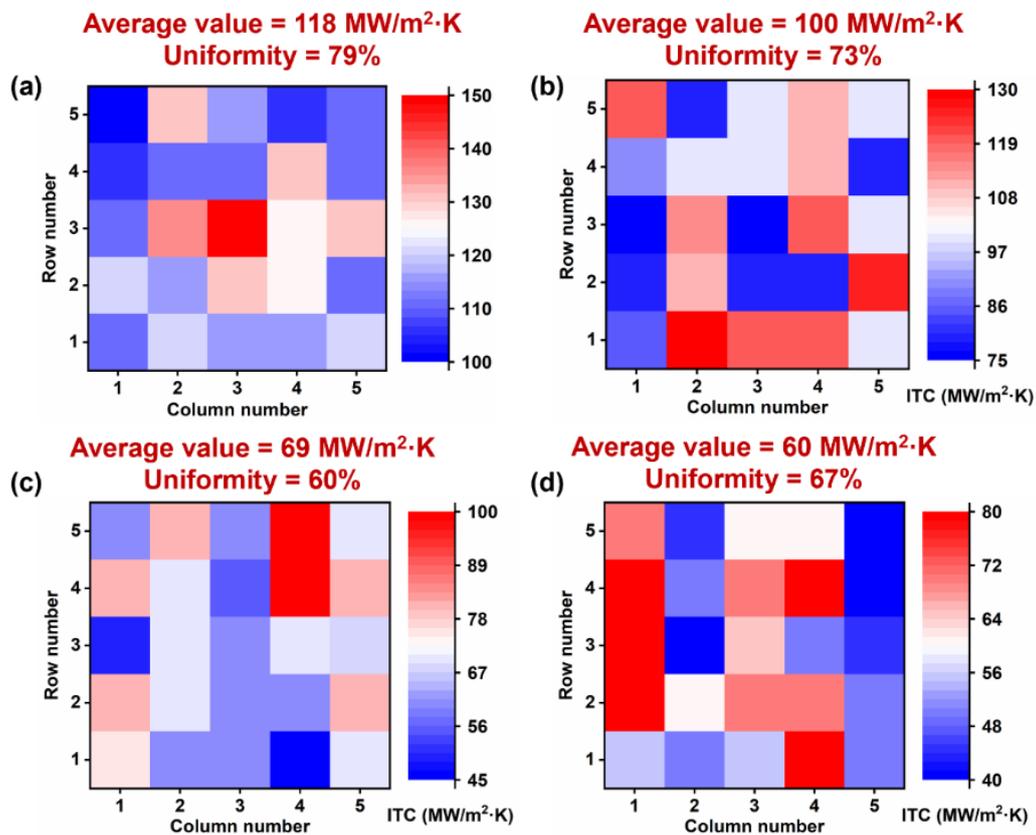

**Fig. S4.** ITC Mapping (5×5 microarray) of (a) the UHV-Ann β-Ga$_2$O$_3$/diamond heterostructure, and the As-TP (b) β-Ga$_2$O$_3$/SiO$_2$/diamond, (c) β-Ga$_2$O$_3$/SiN$_x$/diamond and (d) β-Ga$_2$O$_3$/AlN/diamond heterostructure.

## 4. Scratch adhesion test

The scratch adhesion test is performed using a Rockwell diamond indenter with a 100 μm radius (Fig. S5a). To determine the delamination load, the scratch is performed in ramping mode, starting with a load of zero and increasing to the maximum at the end of the scratch. In this study, 100 μm-long scratches are made, and the load profile along with the lateral movement of the tip is shown in Fig. S5b-c. Adhesion strength is evaluated based on the load of total delamination ($L_{c3}$). The point of delamination is identified from the optical microscope images (Fig. S6).

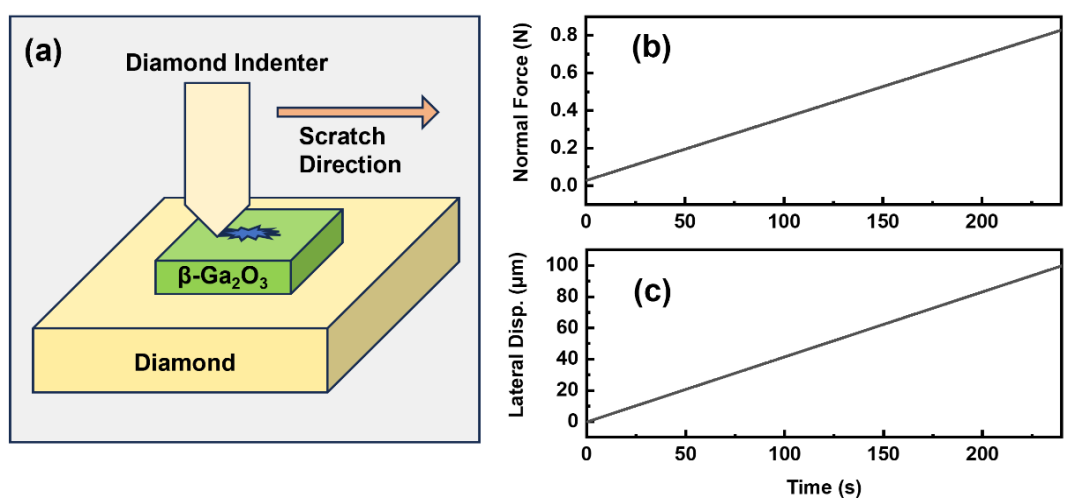

**Fig. S5.** (a) Schematic of scratch test. (b) The load profile (top) and (c) the lateral movement of tip with time during the scratch test.

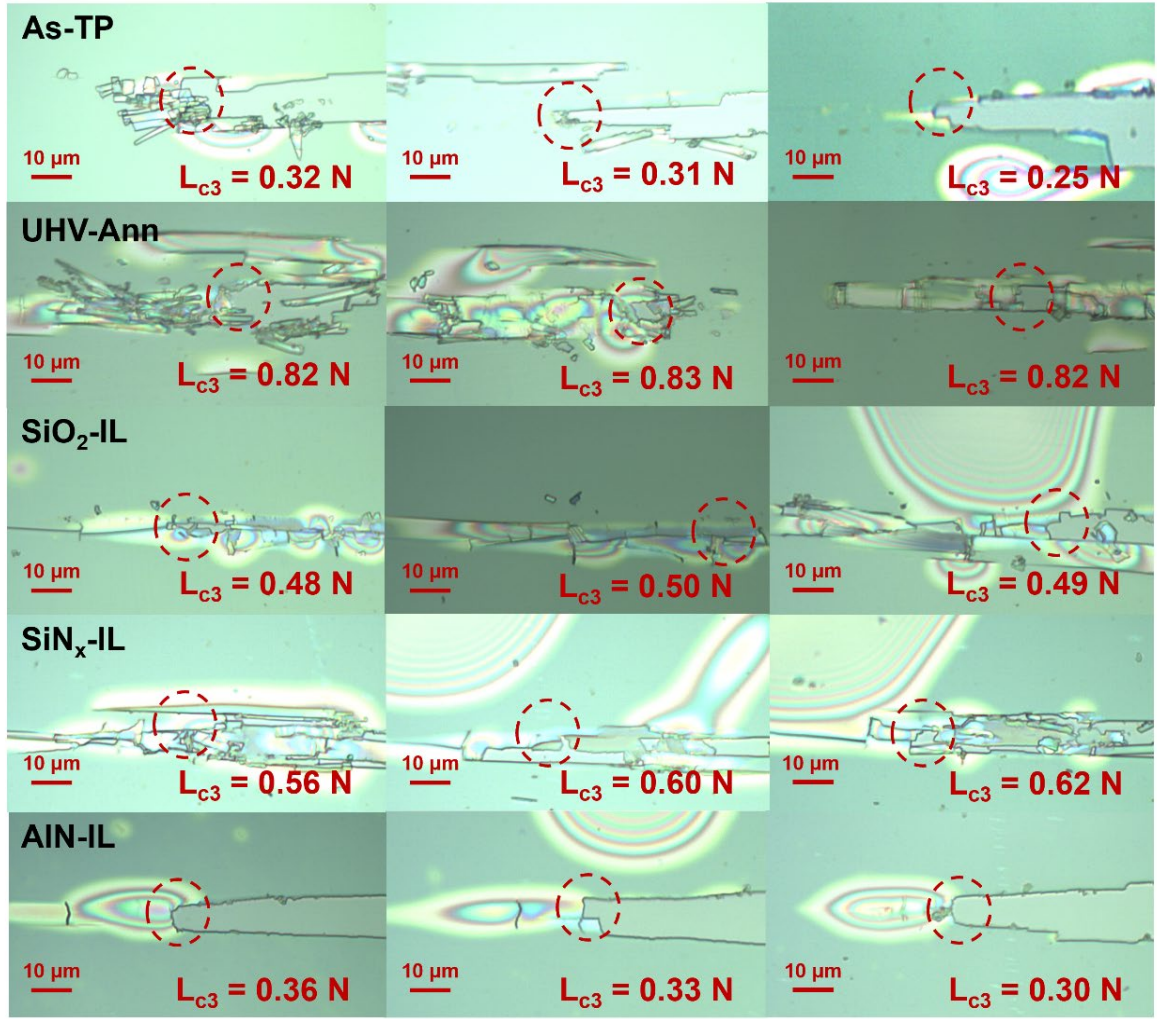

**Fig. S6.** Optical micrograph of the critical scratch failure location.

Given the ultra-thin nature of the β-Ga$_2$O$_3$ film (~250 nm thick) and the extreme hardness of the diamond substrate, the contact deformation during the scratch test is dominated by the elastic response of the substrate rather than the plastic flow of the film[8]. Therefore, the Hertzian contact model is employed to estimate the contact area and interface stresses.

The reduced elastic modulus ($E^*$) of the contact system is primarily governed by the diamond-diamond interaction (indenter and substrate) and is calculated as:

$$\frac{1}{E^*} = \frac{1-v_i^2}{E_i} + \frac{1-v_s^2}{E_s}$$

where $E$ and $v$ represent Young's modulus and Poisson' ratio, respectively. The subscripts i and s denote the diamond indenter and the diamond substrate. Using $E = 1141$ GPa and $v = 0.07$ for diamond[9], the reduced modulus $E^*$ is determined to be ~573 GPa.

The radius of the elastic contact circle ($a$) at the measured critical load ($L_c$) is determined by:

$$a = \left(\frac{3L_c R}{4E^*}\right)^{1/3}$$

where $R$ is the tip radius (100 μm).

The average interface shear strength ($\tau$) is derived by introducing the friction coefficient ($\mu$, 0.15)[10]:

$$\tau = \frac{\mu L_c}{\pi a^2}$$

To quantify the fracture resistance of the interface, the shear strength is converted into adhesion energy ($G_c$)[11], also known as the critical strain energy release rate. This calculation relies on the elastic properties and geometry of the β-$Ga_2O_3$ film:

$$G_c = \frac{\tau^2 h}{2 G_{film}}$$

where $h$ is the film thickness. $G_{film}$ is the shear modulus of the β-$Ga_2O_3$ film, calculated from its Young's modulus ($E_f$, ~200 GPa) and Poisson's ratio ($v_f$, 0.31)[12]:

$$G_{film} = \frac{E_f}{2(1 + v_f)}$$

Given the brittle nature of the β-$Ga_2O_3$/diamond interface, plastic dissipation was considered negligible. Therefore, the measured $G_c$ was approximated to the thermodynamic work of adhesion ($W_{adh}$)[13,14]. Finally, to estimate the bond strength at the atomic level, $G_c$ was converted to the equivalent molar bond energy ($E_{bond}$), assuming failure occurs at the interface governed by the atomic density of the diamond (100) surface:

$$E_{bond} = \frac{G_c \cdot N_A}{n_{surf}}$$

where $N_A$ is the Avogadro constant ($6.022 \times 10^{23}$ mol$^{-1}$) and $n_{surf}$ is the surface atomic density of diamond (100), calculated as $1.57 \times 10^{19}$ atoms m$^{-2}$.

## 5. EELS

The calculated PDOS (Fig. S7) is in excellent agreement with the experimental EELS, and the main spectral features are well reproduced by the theoretical calculations.

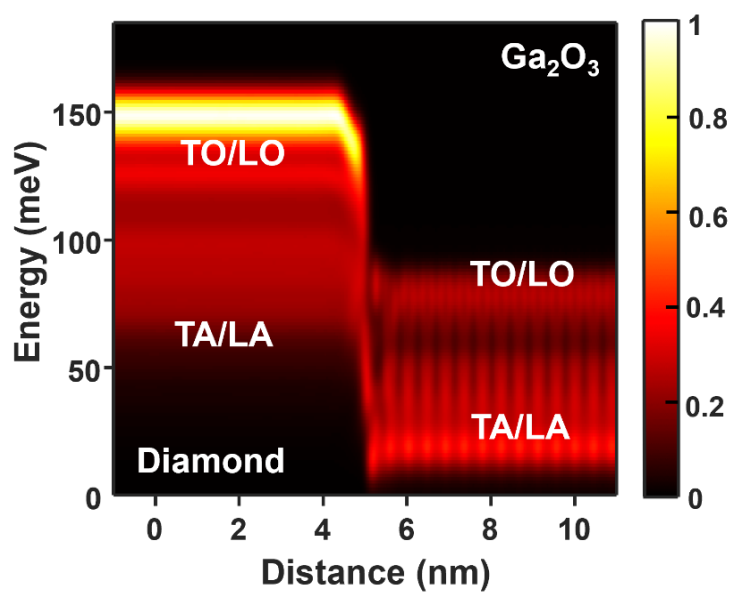

**Fig. S7.** The calculated PDOS of each atom layer.

## 6. MD simulation

As illustrated in Fig. S8, non-equilibrium molecular dynamics is applied for the ITC calculation and spectral analysis. The system is modeled with a combination of potential: a comprehensive core-shell model is used for β-Ga$_2$O$_3$, the Tersoff potential for diamond, and the Lennard-Jones potential for their interface.

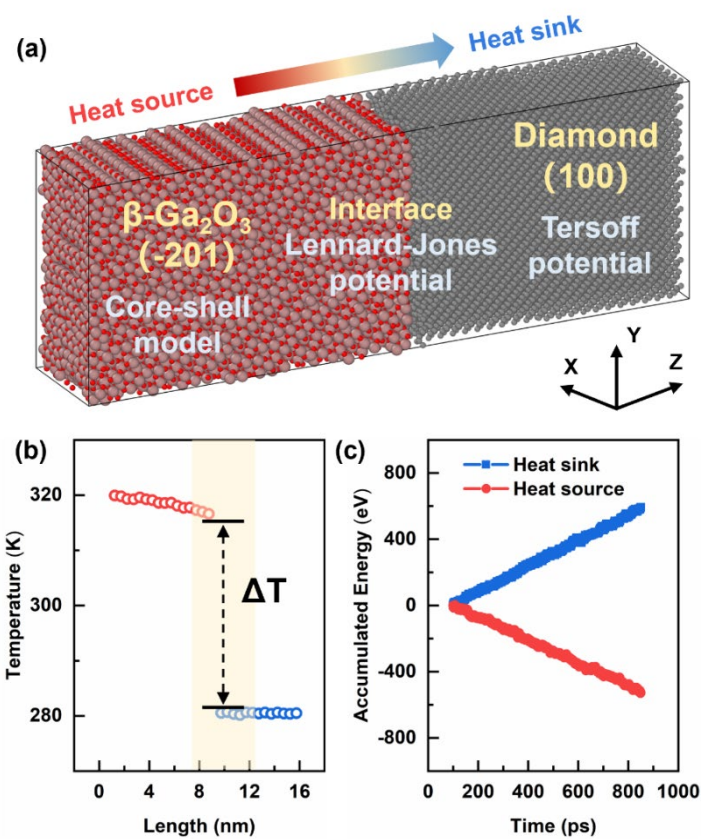

**Fig. S8.** (a) The procedure of diamond-based interface thermal transport calculation and visualized modeling, with (b) the extracted temperature profiles along the z-axis of β-$Ga_2O_3$/diamond interface and (c) variations of accumulated energy as a function of time.

## 7. Sample preparation

The thin films are etched using an inductively coupled plasma (ICP) reactive ion etching (RIE) system (HAASRODE-E200A), with the key parameters detailed in Table S2. The chamber pressure is kept at 10 mTorr, and the substrate holder is cooled to 20°C.

**Table S2.** The etching conditions.

| Sample | Process Gases | Gas Flow Rate (sccm) | ICP Power (W) | RIE Power (W) |
|---|---|---|---|---|
| β-$Ga_2O_3$ | $Cl_2$/$BCl_3$/Ar | 20/50/10 | 500 | 75 |
| Si | CF$_4$/SF$_6$/Ar | 105/11/15 | 300 | 50 |
| GaN | Cl$_2$/BCl$_3$/Ar | 20/50/10 | 500 | 75 |
| LiTaO$_3$ | CHF$_3$/Ar/N$_2$ | 80/20/10 | 500 | 300 |

The release of thin films is achieved by removing the sacrificial layer (SiO$_2$). The samples are immersed in hydrofluoric acid (HF) solution at room temperature (25°C). Samples are etched for different durations based on the characteristics of their respective SiO$_2$ interlayers, with the specific values provided in Table S3.

**Table S3.** The sacrificial layer release parameters.

| Sample | SiO$_2$ Thickness (nm) | Etchant | Etchant Concentration | Etch Time |
|---|---|---|---|---|
| β-Ga$_2$O$_3$/SiO$_2$/Si | 2800 | HF | 40% | ~2 h |
| Si/SiO$_2$/Si | 2000 | HF | 40% | ~8 h |
| GaN/SiO$_2$/Si | 3100 | HF | 40% | ~1 h |
| LiTaO$_3$/SiO$_2$/Si | 4700 | HF | 40% | ~2 h |